\documentclass{elsart}                    

\usepackage{amsmath}
\usepackage{epsf}

\newcommand{\aerr}[2]{\ensuremath{\mathrel{\hbox{\rlap{\hbox{\lower2.4pt\hbox{\scriptsize{#2}}}}\raise4.3pt\hbox{\scriptsize{#1}}}}}}
\newcommand{\aberr}[2]{\ensuremath{\mathrel{\hbox{\rlap{\hbox{\lower4.4pt\hbox{\small{#2}}}}\raise4.4pt\hbox{\small{#1}}}}}}
\newcommand\apm{\aerr{$+$}{$-$}}
\newcommand\abpm{\aberr{$+$}{$-$}}
\newcommand{\xf}{\ensuremath{x_{\!F}}}
\newcommand{\ptsq}{\ensuremath{p_T^2}}
\newcommand{\pt}{\ensuremath{p_T}}
\newcommand{\kt}{\ensuremath{k_t}}
\newcommand{\sqaktsq}{\ensuremath{\sqrt{\langle k_t^2\rangle}}}
\newcommand{\pyth}{\textsc{Py\-thia}}
\newcommand{\pythjet}{\textsc{Py\-thia/Jet\-set}}
\newcommand{\cerenkov}{\v{C}erenkov}
\newcommand{\chidof}{\ensuremath{\chi^2/dof}}
\newcommand{\dn}{\ensuremath{D^{\raise0.3ex\hbox{\scriptsize{\;\!0}}}}}
\newcommand{\dnbar}{\ensuremath{\overline{D}^{\lower0.3ex\hbox{\scriptsize{\;\!0}}}}}
\newcommand{\ccbar}{\ensuremath{c \overline{c}}}
\newcommand{\eg}{e.g.}

\newcommand{\kpi}{\ensuremath{K\pi}}
\newcommand{\kpipipi}{\ensuremath{K\pi\pi\pi}}
\newcommand{\kpic}{\ensuremath{K^-\pi^+}}
\newcommand{\kpipipic}{\ensuremath{K^-\pi^+\pi^-\pi^+}}
\newcommand{\dkpi}{\ensuremath{\dn\!\rightarrow\!K\pi}}
\newcommand{\dkpipipi}{\ensuremath{\dn\!\rightarrow\!K\pi\pi\pi}}
\newcommand{\dkpic}{\ensuremath{\dn\!\rightarrow\!K^-\pi^+}}
\newcommand{\dkpipipic}{\ensuremath{\dn\!\rightarrow\!K^-\pi^+\pi^-\pi^+}}

\def\issue(#1,#2,#3){#1 (#3) #2} 

\def\opcit(#1){ {\em op. cit.}, #1}

\def\ARNPS(#1,#2,#3){Ann.\ Rev.\ Nucl.\ Part.\ Sci.\ \issue(#1,#2,#3)}
\def\CPC(#1,#2,#3){Comp.\ Phys.\ Comm.\ \issue(#1,#2,#3)}
\def\CIP(#1,#2,#3){Comput.\ Phys.\ \issue(#1,#2,#3)}
\def\EPJC(#1,#2,#3){Eur.\ Phys.\ J.\ C\ \issue(#1,#2,#3)}
\def\IEEETNS(#1,#2,#3){IEEE Trans.\ Nucl.\ Sci.\ \issue(#1,#2,#3)}
\def\NP(#1,#2,#3){Nucl.\ Phys.\ \issue(#1,#2,#3)}
\def\NIM(#1,#2,#3){ Nucl.\ Inst.\ and Meth.\ \issue(#1,#2,#3)}
\def\PL(#1,#2,#3){Phys.\ Lett.\ \issue(#1,#2,#3)}
\def\PRD(#1,#2,#3){Phys.\ Rev.\ D \issue(#1,#2,#3)}
\def\PRL(#1,#2,#3){Phys.\ Rev.\ Lett.\ \issue(#1,#2,#3)}
\def\SJNP(#1,#2,#3){Sov.\ J. Nucl.\ Phys.\ \issue(#1,#2,#3)}
\def\ZPC(#1,#2,#3){Zeit.\ Phys.\ C \issue(#1,#2,#3)}

\begin{document}

\begin{flushright}
 FERMILAB-Pub-99/185-E
\end{flushright}
\vspace{1mm}
\begin{frontmatter}

\title{Total Forward and Differential Cross Sections of Neutral $D$ 
	Mesons Produced in 500~GeV/$c$ $\pi^-$--Nucleon Interactions
}
%
%
\collab {Fermilab E791 Collaboration}

\author[umis]{E.~M.~Aitala},
\author[cbpf]{S.~Amato},
\author[cbpf]{J.~C.~Anjos},
\author[fnal]{J.~A.~Appel},
\author[taun]{D.~Ashery},
\author[fnal]{S.~Banerjee},
\author[cbpf]{I.~Bediaga},
\author[umas]{G.~Blaylock},
\author[stev]{S.~B.~Bracker},
\author[stan]{P.~R.~Burchat},
\author[ilit]{R.~A.~Burnstein},
\author[fnal]{T.~Carter},
\author[cbpf]{H.~S.~Carvalho},
\author[usca]{N.~K.~Copty},
\author[umis]{L.~M.~Cremaldi},
\author[yale]{C.~Darling},
\author[fnal]{K.~Denisenko},
\author[ucin]{S.~Devmal},
\author[pueb]{A.~Fernandez},
\author[usca]{G.~F.~Fox},
\author[ucsc]{P.~Gagnon},
\author[cbpf]{C.~Gobel},
\author[umis]{K.~Gounder},
\author[fnal]{A.~M.~Halling},
\author[cine]{G.~Herrera},
\author[taun]{G.~Hurvits},
\author[fnal]{C.~James},
\author[ilit]{P.~A.~Kasper},
\author[fnal]{S.~Kwan},
\author[usca]{D.~C.~Langs},
\author[ucsc]{J.~Leslie},
\author[fnal]{B.~Lundberg},
\author[cbpf]{J.~Magnin},
\author[taun]{S.~MayTal-Beck},
\author[ucin]{B.~Meadows},
\author[cbpf]{J.~R.~T.~de~Mello~Neto},
\author[ksun]{D.~Mihalcea},
\author[tuft]{R.~H.~Milburn},
\author[cbpf]{J.~M.~de~Miranda},
\author[tuft]{A.~Napier},
\author[ksun]{A.~Nguyen},
\author[ucin,pueb]{A.~B.~d'Oliveira},
\author[ucsc]{K.~O'Shaughnessy},
\author[ilit]{K.~C.~Peng},
\author[ucin]{L.~P.~Perera},
\author[usca]{M.~V.~Purohit},
\author[umis]{B.~Quinn},
\author[uwsc]{S.~Radeztsky},
\author[umis]{A.~Rafatian},
\author[ksun]{N.~W.~Reay},
\author[umis]{J.~J.~Reidy},
\author[cbpf]{A.~C.~dos Reis},
\author[ilit]{H.~A.~Rubin},
\author[umis]{D.~A.~Sanders},
\author[ucin]{A.~K.~S.~Santha},
\author[cbpf]{A.~F.~S.~Santoro},
\author[ucin]{A.~J.~Schwartz},
\author[cine,uwsc]{M.~Sheaff},
\author[ksun]{R.~A.~Sidwell},
\author[yale]{A.~J.~Slaughter},
\author[ucin]{M.~D.~Sokoloff},
\author[cbpf]{J.~Solano},
\author[ksun]{N.~R.~Stanton},
\author[fnal]{R.~J.~Stefanski},
\author[uwsc]{K.~Stenson},  
\author[umis]{D.~J.~Summers},
\author[yale]{S.~Takach},
\author[fnal]{K.~Thorne},
\author[ksun]{A.~K.~Tripathi},
\author[uwsc]{S.~Watanabe},
\author[taun]{R.~Weiss-Babai},
\author[prin]{J.~Wiener},
\author[ksun]{N.~Witchey},
\author[yale]{E.~Wolin},
\author[ksun]{S.~M.~Yang},
\author[umis]{D.~Yi},
\author[ksun]{S.~Yoshida},
\author[stan]{R.~Zaliznyak}, and
\author[ksun]{C.~Zhang}

\address[cbpf]{Centro Brasileiro de Pesquisas F{\'\i}sicas, Rio de Janeiro, Brazil}
\address[ucsc]{University of California, Santa Cruz, California 95064, USA}
\address[ucin]{University of Cincinnati, Cincinnati, Ohio 45221, USA}
\address[cine]{CINVESTAV, 07000 Mexico City, DF Mexico}
\address[fnal]{Fermilab, Batavia, Illinois 60510, USA}
\address[ilit]{Illinois Institute of Technology, Chicago, Illinois 60616, USA}
\address[ksun]{Kansas State University, Manhattan, Kansas 66506, USA}
\address[umas]{University of Massachusetts, Amherst, Massachusetts 01003, USA}
\address[umis]{University of Mississippi--Oxford, University, Mississippi 38677, USA}
\address[prin]{Princeton University, Princeton, New Jersey 08544, USA}
\address[pueb]{Universidad Autonoma de Puebla, Mexico}
\address[usca]{University of South Carolina, Columbia, South Carolina 29208, USA}
\address[stan]{Stanford University, Stanford, California 94305, USA}
\address[taun]{Tel Aviv University, Tel Aviv 69978, Israel}
\address[stev]{Box 1290, Enderby, British Columbia V0E 1V0, Canada}
\address[tuft]{Tufts University, Medford, Massachusetts 02155, USA}
\address[uwsc]{University of Wisconsin, Madison, Wisconsin 53706, USA}
\address[yale]{Yale University, New Haven, Connecticut 06511, USA}

\begin{abstract}
We measure the neutral $D$ total forward cross section and the differential cross 
sections as functions of Feynman-$x$ (\xf) and transverse momentum squared for 
500~GeV/$c$\, $\pi^-$--nucleon interactions.  The results are obtained from 
88\,990$\pm$460 reconstructed neutral $D$ mesons from Fermilab experiment E791 
using the decay channels \dkpic\ and \dkpipipic\ (and charge conjugates).  We 
extract fit parameters from the differential cross sections and provide the first 
direct measurement of the turnover point in the \xf\ distribution, 
0.0131$\pm$0.0038.  We measure an absolute \dn+\dnbar\ (\xf$>$0) cross section of 
$15.4 \apm \aerr{1.8}{2.3}$ $\mu$barns/nucleon (assuming a linear $A$ dependence).  
The differential and total forward cross sections are compared to 
theoretical predictions and to results of previous experiments.
\end{abstract}


\end{frontmatter}

Charm hadroproduction is a convolution of short range processes that can be 
calculated in perturbative quantum chromodynamics (QCD) and long range processes 
that cannot be treated perturbatively and thus must be modeled using 
experimental measurements.  The large theoretical uncertainties from both 
contributions are reflected in the relatively large number of 
input parameters that can be adjusted when comparing models to the results of 
experiments.  A single measurement, no matter how precise, cannot 
unambiguously determine these parameters.  However, the results of high 
statistics measurements like the ones reported here, when combined with other 
measurements of similar precision, can constrain such parameters as the charm 
quark mass, the intrinsic transverse momentum of the partons in the incoming hadrons, 
and the effective factorization and renormalization scales used in theoretical 
calculations.

We report here measurements of the differential cross sections versus the 
kinematic variables Feynman-$x$ (\xf) and transverse momentum squared (\ptsq), as 
well as the total forward cross section for the hadroproduction of neutral $D$ 
mesons.  The relatively high pion beam momentum, 500~GeV/$c$, coupled with the good 
geometric acceptance of the Tagged Photon Laboratory (TPL) spectrometer, allows 
us to investigate a wide kinematic region that includes points at negative 
\xf.  We are able to measure the shape of the differential cross section 
versus \xf\ with sufficient precision to confirm, for the first time, that the 
turnover in the cross section does occur at \xf$>$0, as expected for 
incident pions~\cite{pinacoteca}.

Combining data from two \dn\ decay modes, \dkpi\ and \dkpipipi,
\footnote{Charge conjugates are always implied.  We use \dn\ to represent the sum
of \dn\ and \dnbar.  Similarly, \kpi\ (\kpipipi) includes \kpic\ (\kpipipic) 
and charge conjugate.}
we extract a sample of 88\,990$\pm$460 (78730$\pm$430 at \xf$>$0) fully 
reconstructed charm decays to use for these measurements.
In addition to the greater statistical significance, the use of two modes 
provides a means to better understand the systematic errors associated with the
reconstruction of the decay products of these fully-charged decays.

The data were accumulated during the 1991/1992 Fermilab fixed-target run of 
experiment E791~\cite{exp791,e791_pairs}.  The experiment utilized the 
spectrometer built
by the previous TPL experiments, E516~\cite{exp516}, E691~\cite{exp691}, and 
E769~\cite{exp769}, with significant improvements.  The experiment employed a 
500~GeV/$c$ $\pi^-$ beam tracked by eight planes of proportional wire chambers (PWC's) 
and six planes of silicon microstrip detectors (SMD's).  The beam impinged on one 
0.52-mm thick platinum foil (1.6~cm in diameter) followed by four 1.56-mm thick 
diamond foils (1.4~cm in diameter), each foil center separated from the next by 
an average of 1.53~cm, allowing most charm particles to decay in air.  The downstream 
spectrometer consisted of 17 planes of SMD's for vertexing and tracking along 
with 35 planes of drift chambers, 2 PWC planes, and 2 analysis magnets (bending 
in the same direction) for track and momentum measurement.  Two multicell 
threshold 
\cerenkov\ counters, an electromagnetic calorimeter, a hadronic calorimeter, 
and a wall of scintillation counters for muon detection provided particle 
identification.  The trigger was generated using signals from scintillation 
counters as well as the electromagnetic and hadronic calorimeters.  The beam 
scintillation counters included a beam counter 1.3~cm in diameter 
(14~cm upstream of the first
target) and a large beam-halo veto counter with a 1.0~cm hole (8~cm upstream of 
the first target).  The interaction counter was located 2.0~cm downstream of the 
last target and 0.6~cm upstream of the first SMD plane.  The first-level trigger 
required a signal corresponding to at least 1/2 of that expected for a minimum 
ionizing particle 
(MIP) in the beam counter, no signal greater than 1/2 of a MIP in the beam halo 
counter, and a signal corresponding to greater than $\sim$4.5 MIP's in the 
interaction counter (consistent with a hadronic interaction in one of the 
targets).  The second-level trigger required more than 3~GeV of transverse energy 
in the calorimeters.  Additional requirements eliminated events with multiple 
beam particles.
A fast data acquisition system~\cite{e791daq} collected data at rates up to 
30 Mbyte/s with 50~$\mu$s/event deadtime.  Over 2$\times$10$^{10}$ events were 
written to 24\,000 8mm magnetic tapes during a six-month period.

\begin{figure}
  \epsfxsize=\textwidth
  \epsfysize=6.0cm
  \epsffile[5 286 540 546]{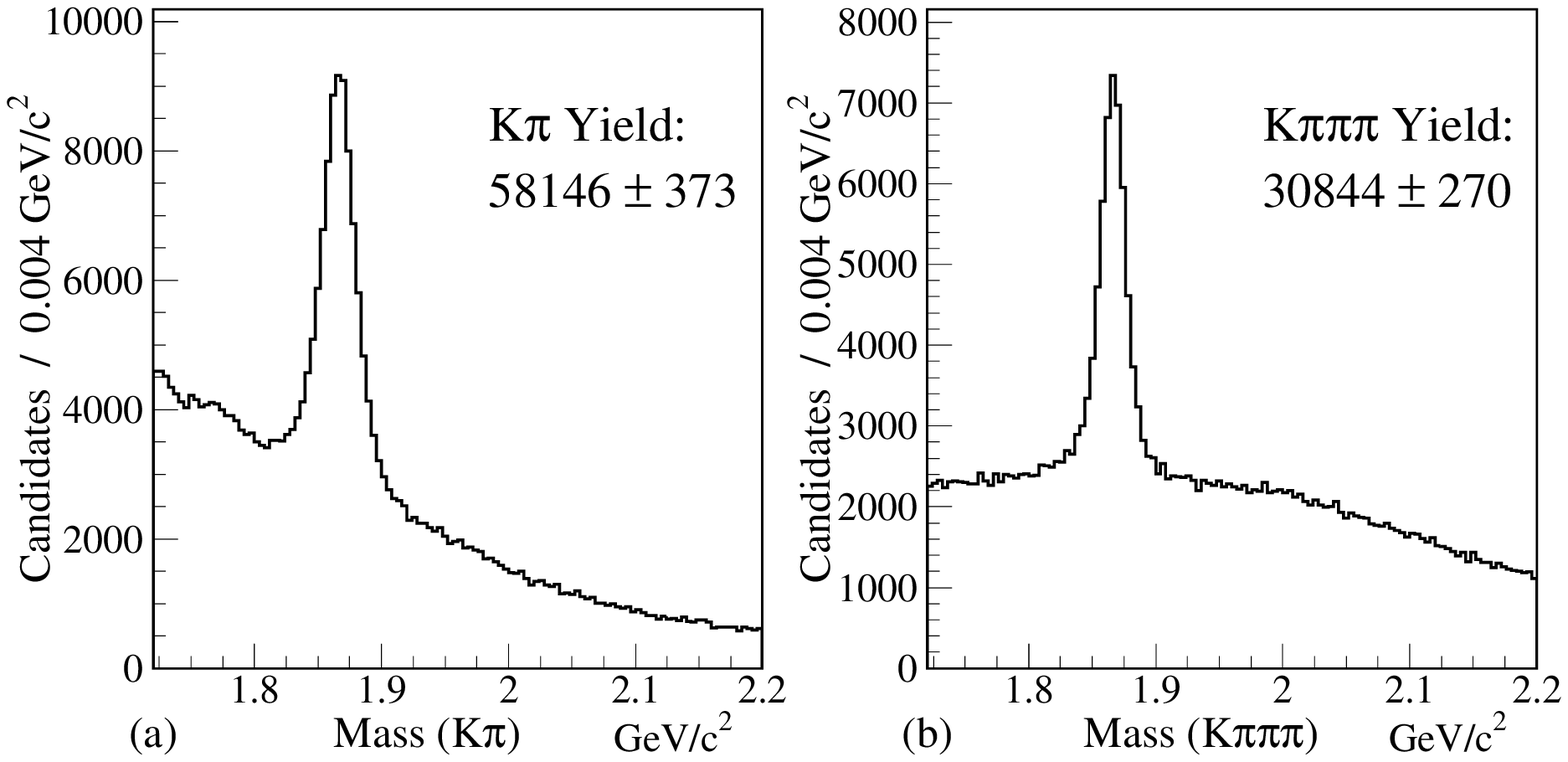}
  \caption{The \kpi\ and \kpipipi\ signals for data with $-0.125$$<$\xf$<$0.8.}
  \label{fig:yield}
\end{figure}

The raw data were reconstructed and filtered~\cite{e791_pairs,farms} to keep 
events with at least two separated vertices, consistent with a primary 
interaction and a charm particle decay.  Following the event reconstruction and 
filtering, selection criteria for the \dn\ 
candidates were determined by maximizing $S / \sqrt{S+B}$ where $S$ is the 
(normalized) number of signal events resulting from a Monte Carlo simulation 
and $B$ is the number of background events appearing in the data sidebands
of the reconstructed \kpi\ or \kpipipi\ mass distribution.  Only selection 
variables that are well modeled by the Monte Carlo simulation were used.  
The final selection criteria varied by decay type (\kpi\ and \kpipipi) and by 
\xf\ region.  The full range of a cut variation is given in the descriptions 
below.   To eliminate 
generic hadronic interaction backgrounds as well as secondary interactions, the 
secondary vertex was required to be longitudinally separated from the primary 
vertex by more than 8-11 times the measurement uncertainty on the longitudinal 
separation 
($\sim$400~$\mu$m) and to lie outside of the target foils.  Backgrounds from the 
primary interaction were also reduced by requiring that the candidate decay 
tracks miss the primary vertex by at least 20-40~$\mu$m.  To ensure a correctly 
reconstructed charm particle and primary vertex, the momentum vector of the \dn\ 
candidate was required to point back to within 35-60~$\mu$m of the primary vertex 
and to have a momentum component perpendicular to the line connecting the 
primary and secondary vertices of less than 350-450~MeV/$c$.  Finally, 
the sum of the squares of the 
transverse momenta of the decay tracks relative to the candidate \dn\ 
momentum vector was required to be greater than 0.4~(GeV/$c$)$^2$
(0.15~(GeV/c)$^2$) for the \kpi\ (\kpipipi) candidates to favor the 
decay of a high-mass particle.
All primary vertices were required to occur in the diamond targets.  Thus, our 
results come from a light, isoscalar target.  The \cerenkov\ information is not 
used in this analysis; all particle-identification combinations are tried.
The inclusive \kpi\ and \kpipipi\ signals are shown in Fig.~\ref{fig:yield}.

The reconstructed data were split into 20 bins of \xf, integrating over all \ptsq, 
and 20 bins of \ptsq, integrating over \xf$>$0.  To 
combine data from varied conditions (\eg, using particles that pass through one and
two magnets), the normalized mass ($m_n$) 
is constructed for each candidate using its calculated mass and error ($m$ and 
$\sigma_m$) and the measured mean mass 
($m_D$): $m_n \equiv \frac{m - m_{D_{\null}}}{\sigma_m}$.  
Using the binned maximum likelihood method, the normalized mass distributions
were fit to a simple Gaussian for the signal and linear or quadratic polynomials
for the background.

\begin{figure}[htbp]
  \epsfxsize=\textwidth
  \epsfysize=6.0cm
  \centerline{\epsffile[0 367 624 735]{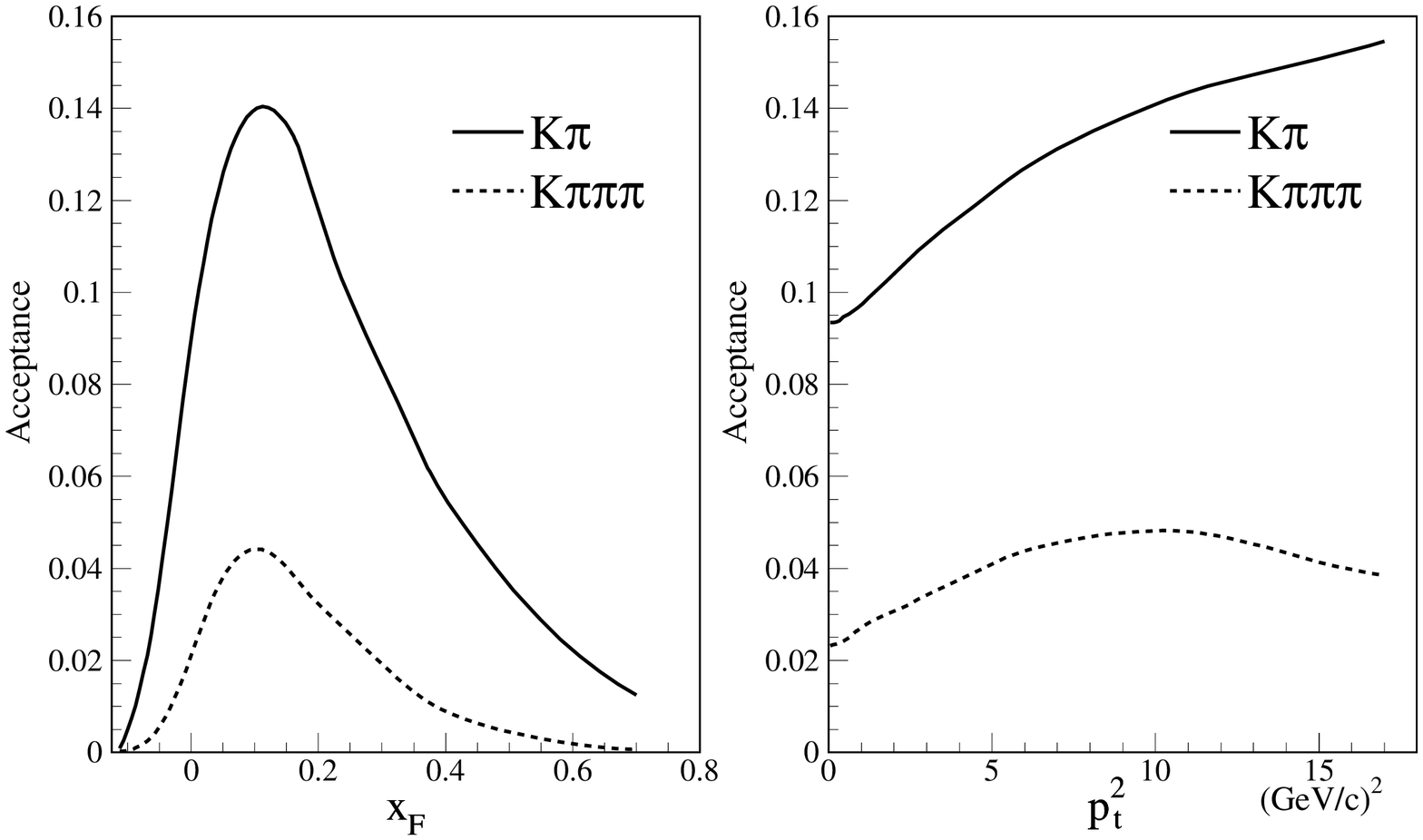}}
  \caption{\dkpi\ and \dkpipipi\ reconstruction efficiencies 
   versus \xf\ (left) and \ptsq\ (right).}
  \label{fig:acc}
\end{figure}

The acceptance can be factorized into trigger efficiency ($\epsilon_{trig}$) and
reconstruction efficiency ($\epsilon_{rec}$).  Most of the trigger inefficiency 
is due to vetoes on multiple beam particles.  These resulted in a 
(70.3$\pm$1.1$\pm$4.2)\% trigger efficiency, where the first error is statistical 
and the second error is systematic.  The interaction and transverse energy 
requirements were greater than 99\% efficient for reconstructable hadronic 
charm decays.  Writing and reading the data tapes was (97.6$\pm$1.0)\% 
efficient.  Combining these efficiencies gives $\epsilon_{trig} = (68.3\pm4.4)\%$, 
where the error is dominated by the systematic error.  The reconstruction 
efficiency is obtained from a Monte Carlo simulation.  The Monte Carlo simulation 
used \pythjet~\cite{pythia} as a physics generator and models the effects of 
resolution, geometry, magnetic fields, and detector efficiencies as well as all 
analysis cuts.  The efficiencies were separately modeled for five evenly spaced 
temporal periods during the experiment.  This was motivated by a highly 
inefficient region of slowly increasing size 
in the center of the drift chambers caused by the 2~MHz pion beam.  The 
Monte Carlo events were weighted to match the observed data distributions of 
\xf, \ptsq, and the summed \ptsq\ of all two-magnet charged tracks in the event 
other than those from the candidate $D$ meson.
The resulting reconstruction efficiencies as a function of \xf\ and \ptsq\ are 
shown in Fig.~\ref{fig:acc}.

\begin{figure}
  \epsfxsize=\textwidth
  \epsfysize=12.0cm
  \epsffile[5 0 631 733]{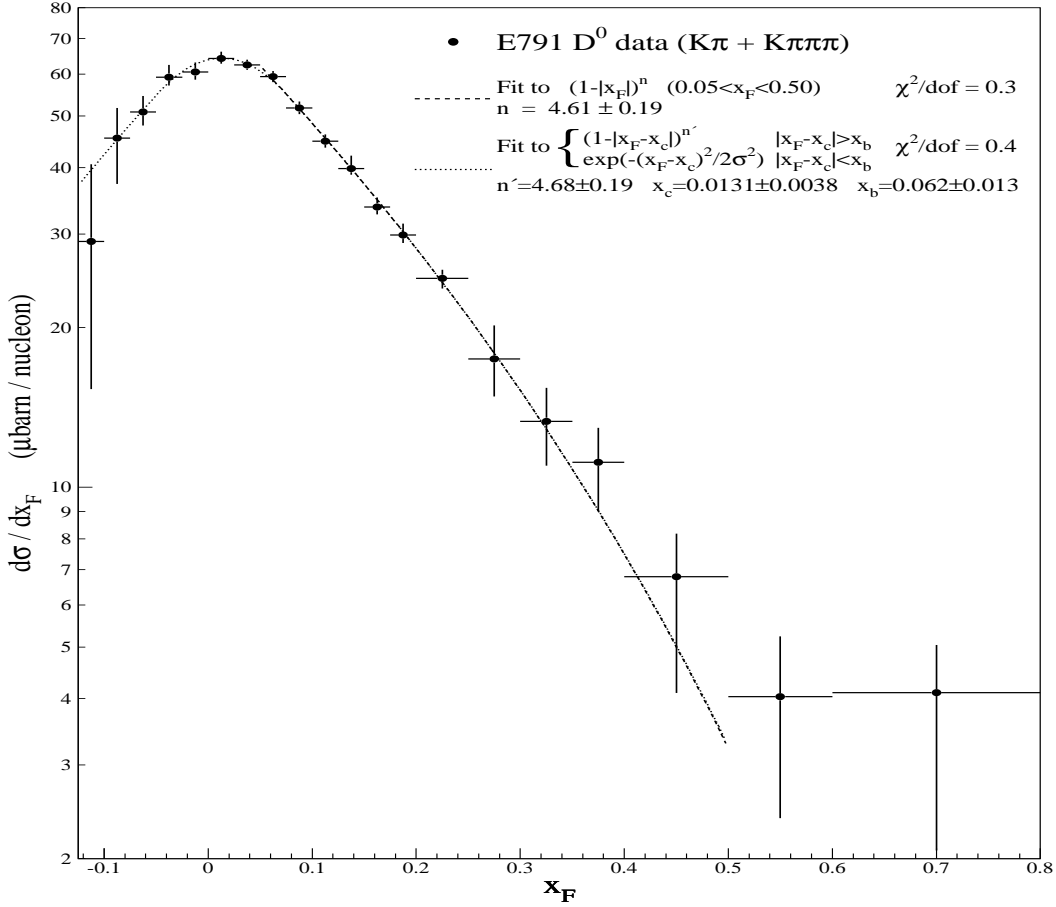}
  \caption{Fits to the \dn+\dnbar\ \xf\ differential cross section
   with the functions given in Eqs.~\ref{eq:xffun} (dashed) and 
   \ref{eq:xfandfun} (dotted).  
   Error bars do not include a $\apm\aerr{10}{11}$\% normalization uncertainty.}
  \label{fig:xf_fit_d0}
\end{figure}

\begin{figure}
  \epsfxsize=\textwidth
  \epsfysize=12.0cm
  \epsffile[5 0 631 733]{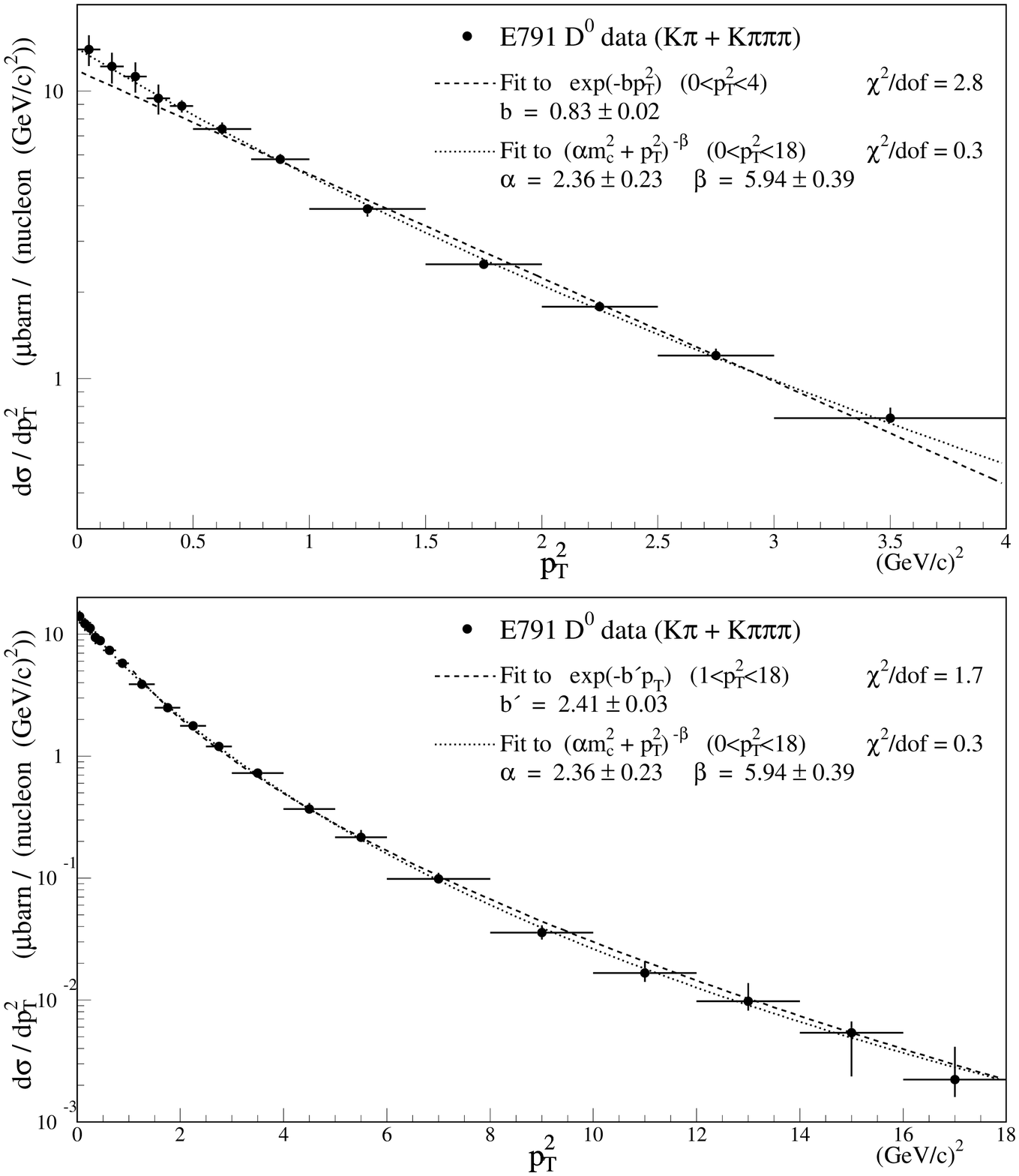}
  \caption{Fits to the \dn+\dnbar\ \ptsq\ differential cross section
   with the functions given in Eqs.~\ref{eq:ptfunb} (dashed, top), 
   \ref{eq:ptfunbp} (dashed, bottom), and \ref{eq:ptfunman} (dotted, top and 
   bottom).  The top plot shows the range 0$<$\ptsq$<$4~(GeV/$c$)$^2$ while the 
   bottom plot shows the full range, 0$<$\ptsq$<$18~(GeV/$c$)$^2$.  
   Error bars do not include a $\apm\aerr{11}{15}$\% normalization uncertainty.}
  \label{fig:pt_fit_d0}
\end{figure}

From the number of reconstructed \dn\ candidates, the reconstruction efficiency,
the trigger efficiency, and the PDG branching fractions~\cite{pdg}, we obtain the 
number of
\dn\ mesons produced in our experiment during the experiment livetime, $N_{prod}$.
The cross section as a function of each variable $z$ (where $z$ = \xf\ or \ptsq), 
is:
\begin{equation}
\sigma(\pi^-N \,\rightarrow\, \dn X ; z) \;\:=\;\: 
\frac{N_{prod}(\dn; z)}{T_N\: N_{\pi^-}}.
\label{eq:sigtot}
\end{equation}
$T_N$ is the number of nucleons per area in the target (calculated from the 
target thickness) and is $(1.224\pm0.004) \times 10^{-6}$ nucleons/$\mu$b.  
$N_{\pi^-}$ is the number of incident $\pi^-$ particles during the experiment 
livetime.  This is obtained directly from a scaler which counted clean beam 
particles ($>$1/2 MIP signals in the beam and interaction counters and no signal 
greater than 1/2 MIP in the beam-halo veto counter) during the experiment 
livetime.  These are the only beam particles which could cause a first-level 
trigger.  Using Eq.~\ref{eq:sigtot}, we obtained the \dn+\dnbar\ differential 
cross sections versus \xf\ and \ptsq\ shown in Figs.~\ref{fig:xf_fit_d0} and 
\ref{fig:pt_fit_d0}.


The systematic errors are divided into two categories and incorporated in two 
stages.  The \emph{uncorrelated} systematic errors are determined individually 
for the \kpi\ and \kpipipi\ results.  These systematic errors include 
uncertainties in the Monte Carlo modeling of the selection criteria, the 
background functions, and the widths used 
in the Gaussian signal functions.  The \emph{correlated} systematic errors are 
calculated for the combined \dn\ result, obtained from adding the \kpi\ and 
\kpipipi\ samples together, weighted by the inverse-square of the combined 
statistical and uncorrelated systematic errors.  The correlated errors 
are associated with uncertainties in the \dn\ lifetime, 
the Monte Carlo production model, the Monte Carlo weighting 
procedure, and the run period weighting procedure.  Finally, we compare our 
measured \kpi\ to \kpipipi\ branching ratio to the PDG~\cite{pdg} value to
estimate the residual tracking and vertexing efficiency modeling error.  In 
addition to these errors, which can affect both the normalization and the shape 
of the differential cross sections, there are two errors which affect only the
normalization: the uncertainties in the trigger efficiency and target thickness.

For the differential cross sections shown in Figs.~\ref{fig:xf_fit_d0} and 
\ref{fig:pt_fit_d0}, the systematic errors are factorized into shape and 
normalization parts.  The error bars in the figures show the sum, in quadrature, 
of the statistical and all systematic errors after factoring out the 
normalization component.  Although the relative importance varies bin-by-bin, 
the most important systematic errors generally come from uncertainties in the 
signal width and the Monte Carlo efficiency modeling.  In all cases, the 
systematic error dominates.  For the total forward cross section, all of the 
errors are summarized and summed in Table~\ref{tab:totsig}.

\begin{table}
\caption{Sources and values of the uncertainties on the 
total forward \dn+\dnbar\ cross section measurement.  The total systematic error comes 
from the quadratic sum of the \dn\ systematic errors.  Although some of the 
systematic errors are subdivided, each \dn\ systematic error is obtained directly
(including correlations) and is not the quadratic sum of the components.}
\vskip 5pt
\label{tab:totsig}
\begin{tabular}{llccc} \hline
\multicolumn{2}{l}{Error type} & \multicolumn{3}{c}{Error (\%)} \\ \hline
& & \kpi\ & \kpipipi\ & \dn\ \\ \hline
\multicolumn{2}{l}{Statistical \& uncorrelated systematic errors} & 
 $\abpm\aberr{4.7}{9.0}$ & $\abpm\aberr{7.3}{11.0}$ & $\abpm\aberr{4.7}{7.6}$ \\
& Statistics  & $\pm$1.0 & $\pm$2.8 & \\
& Selection criteria efficiency modeling & $\abpm\aberr{0.1}{0.0}$ & $\abpm\aberr{1.2}{0.0}$ & \\
& MC background function & $\abpm\aberr{0.0}{5.5}$  & $\abpm\aberr{0.0}{6.0}$ & \\
& \dn\ signal width in fits & $\abpm\aberr{4.0}{4.1}$ & $\abpm\aberr{3.9}{4.4}$ & \\
& PDG~\cite{pdg} branching ratio & $\pm$2.3 & $\pm$5.3 & \\
\multicolumn{2}{l}{Correlated systematic errors} & & & $\abpm\aberr{6.9}{9.9}$ \\
& \dn\ lifetime & \multicolumn{2}{c}{$\abpm\aberr{0.6}{0.5}$} & \\
& Monte Carlo kinematic weighting & \multicolumn{2}{c}{$\abpm\aberr{0.0}{2.3}$} & \\
& Monte Carlo production model & \multicolumn{2}{c}{$\abpm\aberr{5.9}{5.1}$} & \\
& Time dependent efficiency modeling & \multicolumn{2}{c}{$\abpm\aberr{1.0}{2.0}$}  & \\
\multicolumn{2}{l}{Tracking and vertex finding} &
& & $\pm$4.6 \\
\multicolumn{2}{l}{Trigger efficiency} & 
& & $\pm$6.4 \\
\multicolumn{2}{l}{Target thickness} & 
& & $\pm$0.3 \\ \hline
\multicolumn{2}{l}{Total} & 
& & $\abpm\aberr{11.5}{14.8}$ \\ \hline
\end{tabular}
\end{table}

In the past, \xf\ distributions have been fit with 
\begin{equation}
\frac{d \sigma}{d \xf} \;\:=\;\: A (1 - |\xf|)^n .
\label{eq:xffun}
\end{equation}
Fitting Eq.~\ref{eq:xffun} in the range $0.05\!<\!\xf\!<\!0.50$, we find 
$n=4.61\pm0.19$, as shown in Fig.~\ref{fig:xf_fit_d0}.  This function 
does not provide a complete representation of our data.  Although the \chidof\ 
is small 
(0.3), the value of $n$ is quite dependent on the range fitted and on the errors 
on the data points.  Another function, which can be extended into the negative 
\xf\ region, is an extension of Eq.~\ref{eq:xffun} which uses a power-law
function in the tail region and a Gaussian in the central region; that is,
\begin{equation}
\frac{d \sigma}{d \xf} \;\:=\;\: 
\begin{cases}
A (1 - |\xf - x_c|)^{n'} , & |\xf - x_{c_{\null}}| > x_b \\
A' \exp{\left[-\frac{1}{2}(\frac{\xf - x_c}{\sigma})^2\right]} , & |\xf - x_c| < x_b
\end{cases}
.
\label{eq:xfandfun}
\end{equation}
Requiring continuous functions and derivatives allows us to write 
Eq.~\ref{eq:xfandfun} with one normalization parameter and three shape 
parameters: $n'$ gives the shape in the tail region, $x_c$ is the 
turnover point, and $x_b$ is the boundary between the Gaussian and power-law 
function.  The fit parameters from this function are nearly independent of
the fit range.  Fitting our data in the range $-0.125$$<$\xf$<$0.50 
gives $n'=4.68\pm0.21$, $x_c=0.0131\pm0.0038$, and $x_b=0.062\pm0.013$ with 
a \chidof=0.4, as shown in Fig.~\ref{fig:xf_fit_d0}.  This is the first 
measurement of the turnover point $x_c$ in the charm sector.  The fact that
it is significantly greater than zero is consistent with a harder gluon 
distribution in the beam pions than in the target nucleons.

The functions which have been used in the past to fit the \ptsq\ distribution 
are:
\begin{equation}
\frac{d \sigma}{d \ptsq} \;\:=\;\: A e^{-b \ptsq}
\label{eq:ptfunb}
\end{equation}
at low \ptsq\ (\ptsq$\,<\,$4.0~(GeV/$c$)$^2$ for this analysis), 
\begin{equation}
\frac{d \sigma}{d \ptsq} \;\:=\;\: A e^{-b' \pt}
\label{eq:ptfunbp}
\end{equation}
at high \ptsq\ (\ptsq$\,>\,$1.0~(GeV/$c$)$^2$ for this analysis), and 
\begin{equation}
\frac{d \sigma}{d \ptsq} \;\:=\;\: \left[\frac{A}{\alpha\,m_c^2 \:+\: \ptsq} \right]^{\beta}
\label{eq:ptfunman}
\end{equation}
over all \ptsq\ with $m_c$ set to 1.5~GeV/$c$$^2$~\cite{frixione}.
The results of fitting these equations to the data are shown in 
Fig.~\ref{fig:pt_fit_d0}.  For the ranges given above, the fit results are:
\begin{itemize}
\item{$b$ = 0.83$\pm$0.02 with \chidof\ = 2.8,}
\item{$b'$ = 2.41$\pm$0.03 with \chidof\ = 1.7, and}
\item{$\alpha$ = 2.36$\pm$0.23~(GeV/$c$$^2$)$^{-2}$ and $\beta$ = 5.94$\pm$0.39 
with \chidof\ = 0.3.}
\end{itemize}

Equation~\ref{eq:ptfunb} does not provide a good fit even over the very limited 
range to which it is applied.  While the \chidof\ (1.7) of the fit to 
Eq.~\ref{eq:ptfunbp} is not good, it appears to be a reasonable fit to the 
data.  Equation~\ref{eq:ptfunman} provides a very good fit to the data over the 
entire range of \ptsq\@.  Unfortunately, using two free parameters (in addition 
to the normalization) makes it more difficult to compare to other experiments and 
theory since the parameters in this fit are highly correlated.  
This is reflected in the large (7-10\%) errors on $\alpha$ and $\beta$ compared 
to the error on $b'$ (1\%), as shown above.

\begin{figure}
  \epsfxsize=\textwidth
  \epsfysize=12.0cm
  \epsffile[4 0 632 736]{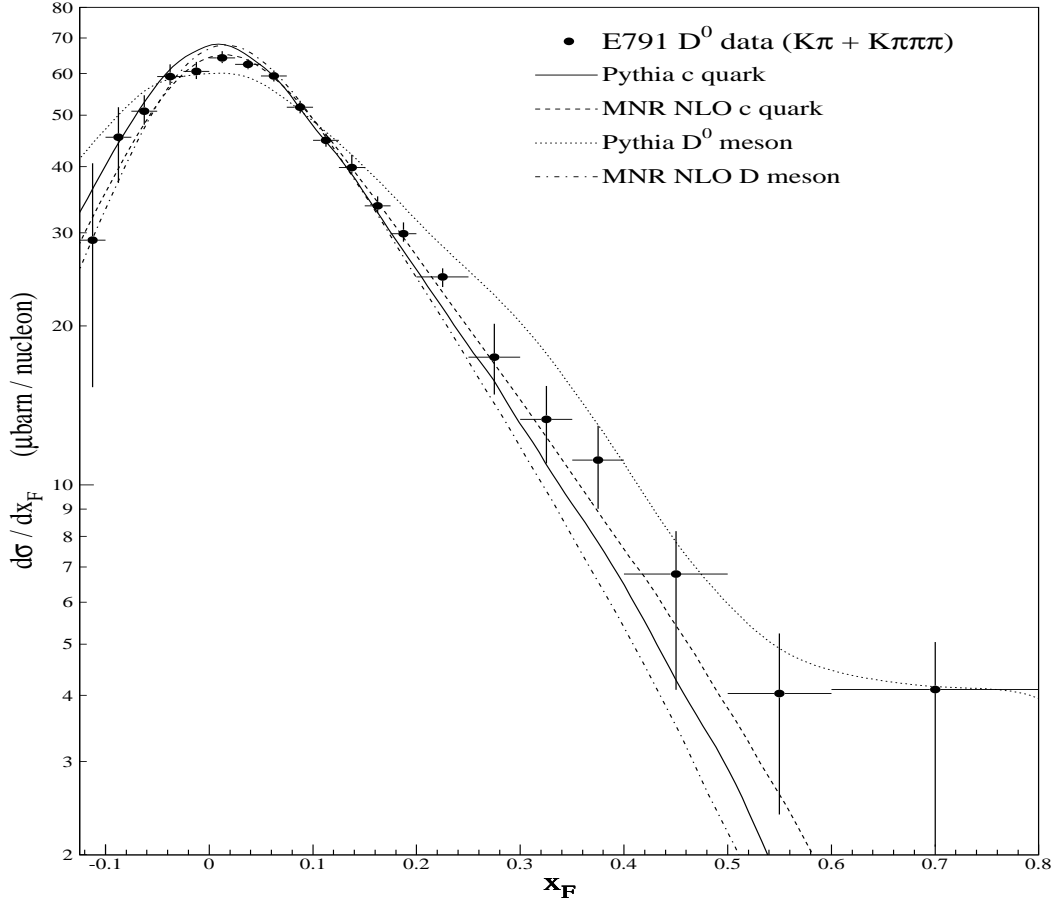}
  \caption{The \dn+\dnbar\ \xf\ differential cross section compared to 
   various theoretical predictions described in the text.  The curves
   are normalized to obtain the best fit to the data in each case.  
   Error bars do not include a $\apm\aerr{10}{11}$\% normalization uncertainty.}
  \label{fig:xf_theory_d0}
\end{figure}

\begin{figure}
  \epsfxsize=\textwidth
  \epsfysize=12.0cm
  \epsffile[0 2 630 731]{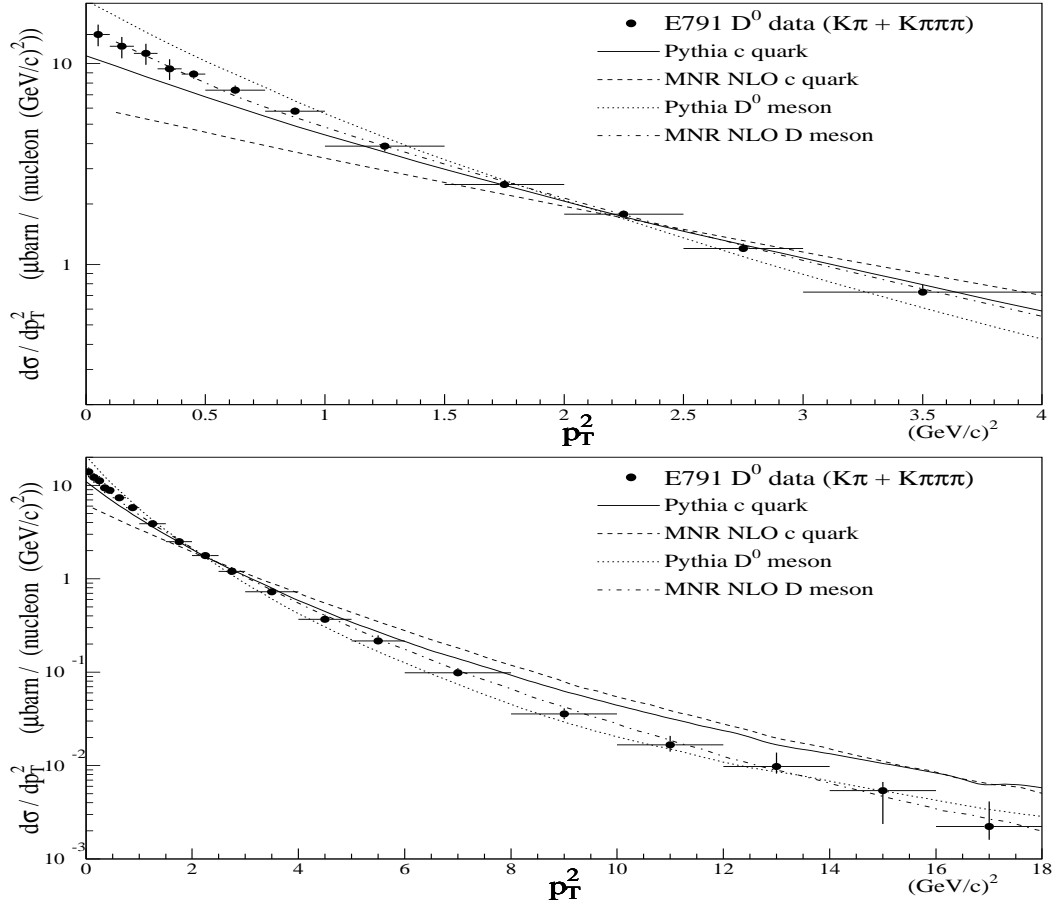}
  \caption{The \dn+\dnbar\ \ptsq\ differential cross section compared to 
   various theoretical predictions described in the text.  The curves
   are normalized to obtain the best fit to the data in each case.  The top plot 
   shows the range 0$<$\ptsq$<$4~(GeV/$c$)$^2$ while the bottom plot shows the full 
   range, 0$<$\ptsq$<$18~(GeV/$c$)$^2$.  Error bars do not include a 
   $\apm\aerr{11}{15}$\% normalization uncertainty.}
  \label{fig:pt_theory_d0}
\end{figure}

Figures~\ref{fig:xf_theory_d0} and \ref{fig:pt_theory_d0} show a comparison of 
our \xf\ and \ptsq\ distributions to theoretical predictions for charm quark and 
$D$ meson production.  Although the data come from \dn\ mesons, the theoretical
predictions for charm quark production are included for completeness. 
The theoretical curves are normalized to obtain the best fit (lowest \chidof) to 
our data.  The theoretical predictions come from a next-to-leading order (NLO) 
calculation by Mangano, Nason, and Ridolfi (MNR)~\cite{mnrtheory} and the 
\pythjet~\cite{pythia} event generator.  The MNR NLO charm quark calculation uses 
SMRS2~\cite{pdf_smrs} (HMRSB~\cite{pdf_hmrsb}) NLO parton distribution functions 
for the pion (nucleon), a charm quark mass of 1.5~GeV/$c$$^2$, and an average 
intrinsic transverse momentum of the incoming partons (\sqaktsq) of 1.0~GeV/$c$.
The value for \sqaktsq\ was suggested by M.~L.~Mangano~\cite{mlmsug} and is 
independently motivated by the study of azimuthal angle correlations between two 
charm particles in the same event~\cite{frixione}.
The $D$ meson results are obtained by convoluting the charm quark results with
the Peterson fragmentation function~\cite{peterson} with $\epsilon = 0.01$.
The low value for $\epsilon$ was also suggested by M.~L.~Mangano~\cite{mlmsug} in 
response to a reanalysis of $D$ fragmentation in $e^+e^-$ 
collisions~\cite{dfragee}.  The \pythjet\ event generator uses leading order 
DO2~\cite{pdf_do2} (CTEQ2L~\cite{pdf_cteq2}) parton distribution functions for 
the pion (nucleon), a charm quark mass of 1.35~GeV/$c$$^2$, \sqaktsq\ of 
0.44~GeV/$c$, and the Lund string fragmentation scheme to obtain \dn\ 
results.  Tables~\ref{tab:comp_xf} and \ref{tab:comp_pt} show a comparison of our 
\xf\ and \ptsq\ fit results to theoretical predictions and to recent high-statistics 
charm experiments which used pion beams.  The evident energy dependence of the 
shape parameters in Tables~\ref{tab:comp_xf} and \ref{tab:comp_pt} are consistent
with theoretical predictions~\cite{frixione}.


\begin{table}
\caption{Comparison of \xf\ shape parameters to recent high-statistics pion-beam 
  charm production experiments and to theory.   E791 results are for \dn\ mesons; 
  E769 (WA92) results are from a combined sample of \dn, $D^+$, and $D_s$ 
  (\dn\ and $D^+$) mesons.  The theoretical results are obtained using the default 
  parameters (described in the text).  The fit range for $n$ (Eq.~\ref{eq:xffun}) 
  is in the table.  The fit range for $n'$ and $x_c$ (Eq.~\ref{eq:xfandfun}) is 
  $-0.125$$<$\xf$<$0.50.}
\vskip 5pt
\label{tab:comp_xf}
\begin{tabular}{lccccc} \hline
Experi- & Energy & \xf\ Range$\!$ & $n$ & $n'$ &  $x_c$            \\
ment    & (GeV)  &            &     &      &                   \\ \hline
E791              & 500 & 0.05--0.5 & $\!$4.61$\pm$0.19 & 4.68$\pm$0.21 & .0131$\pm$.0038 \\
WA92\cite{wa92_1} & 350 & 0.0--0.8  & 4.27$\pm$0.11     &                       & \\
E769\cite{e769_2} & 250 & 0.0--0.8  & 4.03$\pm$0.18     &                       & \\
MNR NLO $c$ & 500 & 0.05--0.5  & 4.68 & 5.06 & 0.0231 \\
MNR NLO $D$ & 500 & 0.05--0.5  & 5.53 & 6.00 & 0.0237 \\
Pythia $c$ & 500 & 0.05--0.5  & 5.01 & 5.12 & 0.0115  \\
Pythia \dn\ & 500 & 0.05--0.5  & 3.62 & 3.66 & 0.0041 \\
\hline
\end{tabular}
\end{table}

\begin{table}
\caption{Comparison of \ptsq\ shape parameters to recent high-statistics 
  pion-beam charm production experiments and to theory.  
  Data samples are as described in Table~\ref{tab:comp_xf}.
  The fit range for $b$ (Eq.~\ref{eq:ptfunb}) is 0$<$\ptsq$<$4~(GeV/$c$)$^2$ except
  for WA92 which is 0$<$\ptsq$<$7~(GeV/$c$)$^2$\@.
  The functions used to extract $b'$ (Eq.~\ref{eq:ptfunbp}) and $\alpha, \beta$ 
  (Eq.~\ref{eq:ptfunman}) are fit in the range \ptsq$>$1~(GeV/$c$)$^2$ and 
  \ptsq$<$18~(GeV/$c$)$^2$, respectively.}
\vskip 5pt
\label{tab:comp_pt}
\begin{tabular}{lcccccc} \hline
Experi- & Energy & $b$  & $b'$           &  $\alpha$ & $\beta$        \\
ment & $\!$(GeV) & $\!\!$(GeV/$c$)$^2\!$ & (GeV/$c$)$^{-2}$ & $\!$(GeV/$c$)$^{-1}\!$ & & \\ \hline
E791              & 500 & 0.83$\pm$0.02 & 2.41$\pm$0.03 & 2.36$\pm$0.23 & 5.94$\pm$0.39 \\
WA92\cite{wa92_1}$\!$ & 350 & 0.89$\pm$0.02  &    &              & \\
E769\cite{e769_2} & 250 & 1.08$\pm$0.05 & 2.74$\pm$0.09 & 1.4$\pm$0.3 & 5.0$\pm$0.6 \\
MNR NLO $c$ & 500 & 0.57 & 1.88 & 6.20 & 8.68 \\
MNR NLO $D$ & 500 & 0.94 & 2.32 & 1.99 & 5.30 \\
Pythia $c$  & 500 & 0.77 & 2.09 & 2.32 & 5.14 \\
Pythia \dn & 500 & 1.06 & 2.58 & 1.55 & 5.07 \\
\hline
\end{tabular}
\end{table}

\begin{figure}
  \epsfxsize=\textwidth
  \epsfysize=10.0cm
  \epsffile[17 354 506 736]{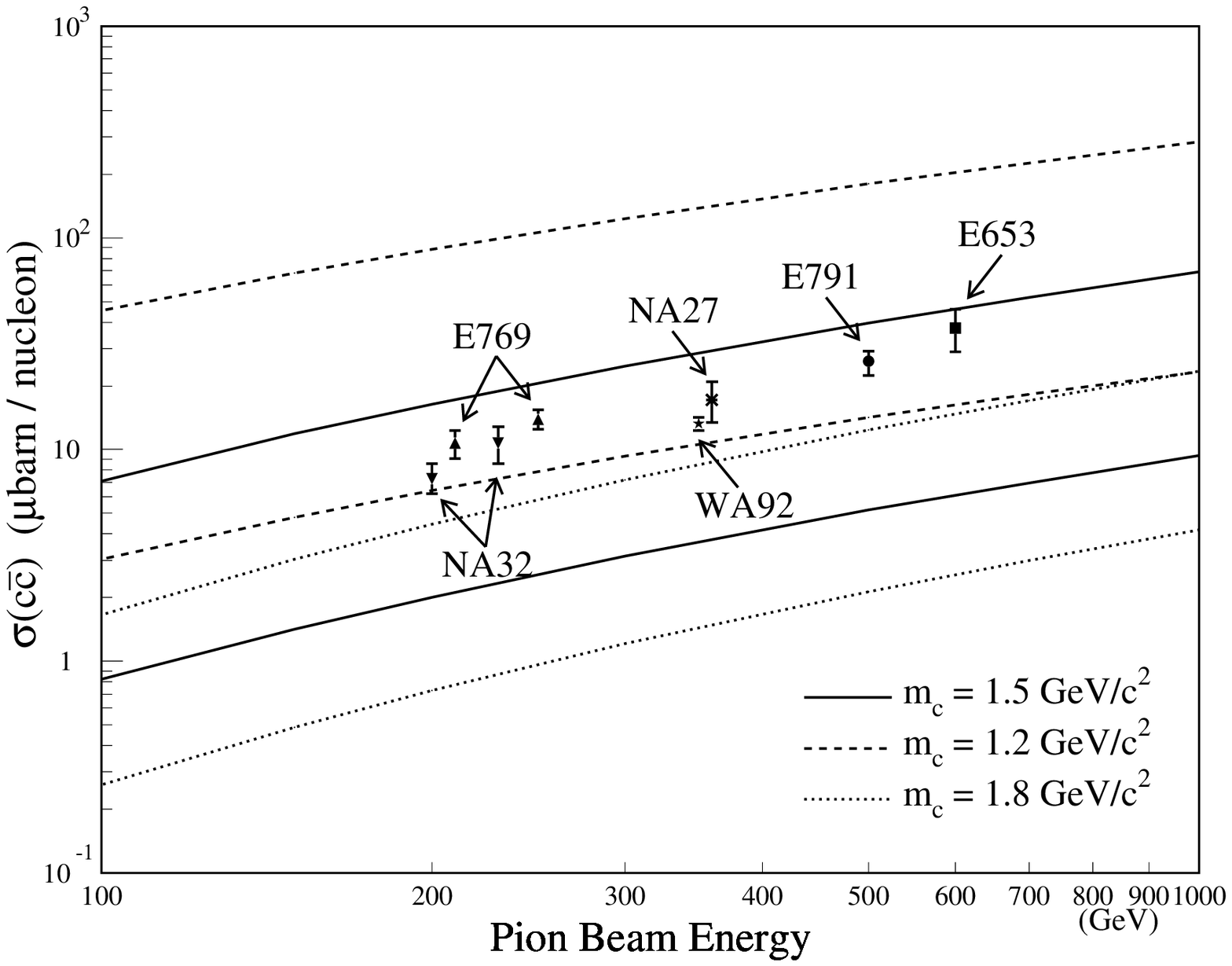}
  \caption{Theoretical and experimental results for the $\pi$--N $\rightarrow$ 
   \ccbar\ cross section
   versus energy.  The theoretical curves are obtained from MNR NLO 
   predictions~\cite{fmnr_hqp97}.  The three bands (solid, dashed, and dotted) 
   correspond to three different charm quark masses (1.5, 1.2, 
   and 1.8~GeV/$c$$^2$).  The variation within the bands comes only from varying 
   the renormalization scale.  The factorization scale is kept fixed.
   All experimental data points are 1.7 times the single inclusive 
   \dn+\dnbar\ \xf$>$0 cross sections.  The E653, NA27, WA92, NA32 (200~GeV), 
   NA32 (230~GeV), and E769 data were obtained from references \cite{e653_1}, 
   \cite{na27_1}, \cite{wa92_1}, \cite{na32_1}, \cite{na32_2}, and \cite{e769_1}, 
   respectively.}
  \label{fig:mnretot}
\end{figure}

We obtain the total forward cross section by summing the \xf\ differential cross 
section for \xf$>$0 and assuming the cross section for 0.8$<$\xf$<$1.0 is half 
that of the cross section for 0.6$<$\xf$<$0.8 but with the same error.  Assuming 
a linear dependence on the atomic number~\cite{e769_adep}, we obtain the neutral 
$D$ total forward cross section,
$\sigma(\dn\!+\dnbar;\xf\!>0) \,=\, 15.4 \apm \aerr{1.8}{2.3}$ $\mu$barns/nucleon.
To obtain the total charm cross section, $\sigma(\ccbar)$, we multiply our 
\dn+\dnbar cross section by 1.7.  This accounts for three multiplicative effects:
the relative production of 
charm quarks compared to \dn\ mesons (2.1), the conversion from \xf$>$0 to all 
\xf\ (1.6), and the conversion to the \ccbar\ cross section from the sum of charm plus 
anticharm cross sections (0.5)~\cite{fmnr_hqp97}. 
We compare our total charm cross section to other experiments and to the NLO 
predictions as a function of pion-beam energy in Fig.~\ref{fig:mnretot}.  All 
experimental results are obtained by multiplying the \dn+\dnbar cross section 
by 1.7.  The rise of the charm production cross section with energy is modeled
reasonably well by the NLO theory, although the absolute value at any
point depends greatly on the input parameters to the theory.

In this paper we have presented the total forward cross section and differential 
cross sections versus \xf\ and \ptsq\ for \dn\ mesons from Fermilab experiment 
E791 data.  This analysis represents the first measurement of the \dn\ cross 
section for a 500~GeV/$c$ pion beam.  The high statistics allows us to clearly 
observe a turnover point greater than zero ($x_c$ = 0.0131$\pm$0.0038) in the 
Feynman-$x$ distribution, providing evidence for a harder gluon distribution in 
the pion than in the nucleon.

We have compared our differential cross section results to predictions from the 
next-to-leading order calculation by Mangano, Nason, and Ridolfi~\cite{mnrtheory} 
and to the Monte Carlo event generator \pyth\ by 
T.~Sj\"{o}strand~\etal~\cite{pythia}.  With suitable choices for the intrinsic 
\kt\ of the partons and the Peterson fragmentation function parameter, the 
NLO $D$ meson calculation provides a good match to the \ptsq\ spectra and
a fair match to the \xf\ distribution.  The string fragmentation scheme in 
\pyth\ softens the original charm quark \ptsq\ distribution too much, and 
hardens the \xf\ spectra too much in both directions.  However, the \dn\ result 
does predict the flattening of the \xf\ cross section at high \xf.  
The many adjustable parameters in the 
theoretical models allow one to obtain distributions which are quite consistent 
with these data.  Unfortunately, a given set of parameters is neither unique, nor 
does it necessarily provide a good match to other data.  In conjunction with 
other charm production results from this and other recent high-statistics 
experiments, however, it may be
possible to find a unique set of parameters.  These results come 
from experiments with a variety of beam energies and types, and include 
measurements of differential cross sections~\cite{wa92_1,e769_1,e781},
production asymmetries~\cite{wa92_1,e769_1,e791_asym,e769_asym,e781,e687_asym}, 
and correlations between two charm particles in the same 
event~\cite{e791_pairs,wa92_pairs,e687_pairs}.

Unlike the uncertainties in the theoretical calculations of the differential 
cross sections, the uncertainties in the theoretical calculation of the total 
cross section come mostly from the perturbative calculation.  The relatively 
large uncertainties are due to the low mass of the charm quark, which results 
in a large (unknown) contribution from higher-order terms.  The total forward 
\dn+\dnbar\ cross section measured by E791 is 
$\sigma(\dn\!+\dnbar\;;\xf\!>0) \,=\, 15.4 \apm \aerr{1.8}{2.3}$ $\mu$barns/nucleon, 
assuming a linear atomic number dependence.  The cross section is consistent with 
the MNR NLO prediction.

We express our special thanks to S.~Frixione, M.~L.~Mangano, P.~Nason, and 
G.~Ridolfi for the use, and help in the use, of their NLO QCD software.  
We gratefully acknowledge the assistance from Fermilab and other
participating institutions.  This work was supported by the Brazilian Conselho
Nacional de Desenvolvimento Cient\'\i fico e Technol\'{o}gico, CONACyT
(Mexico), the Israeli Academy of Sciences and Humanities, the U.S. Department
of Energy, the U.S.-Israel Binational Science Foundation, and the U.S.
National Science Foundation.

\bibliographystyle{unsrt}

\end{document}